\documentclass[preprint2]{aastex}

\def\>{$>$}
\def\<{$<$}
\def\sun{$_{\odot}$}

\def\nh{N$_{\rm H}$ }

\def\xb1916{XB 1916-053}

\def\sa{S$_a$~}
\def\Mdot{\hbox{$\dot {M}$}}

\begin{document}


\title{X-ray Dip Monitoring of XB 1916-053}


\author{T. Narita}
\affil{Department of Physics, College of the Holy Cross,
    Worcester, MA 01610}
%\email{tnarita@holycross.edu}

\author{J. E. Grindlay}
\affil{Harvard-Smithsonian Center for Astrophysics, Cambridge, MA 02138}
%\email{aastex-help@aas.org}

\author{P. F. Bloser}
\affil{NASA/Goddard Space Flight Center, Greenbelt, MD 20771 }
\and
\author{Y. Chou}
\affil{National Tsing Hua University, Taiwan}

\begin{abstract}
We report on the long term monitoring of X-ray dips from the
ultracompact low-mass X-ray binary (LMXB) XB 1916-053.  Roughly
one-month interval observations were carried out with the {\it Rossi
X-ray Timing Explorer} (RXTE) during 1996, during which the source
varied between dim, hard states and more luminous, soft states. The
dip spectra and dip lightcurves were compared against both the
broadband luminosity and the derived mass accretion rate (\Mdot). The dips
spectra could be fitted by an absorbed blackbody plus cut-off power
law non-dip spectral model, with additional absorption ranging from 0
to $>100\times10^{22}$ cm$^{-2}$.  The amount of additional blackbody
absorption was found to vary with the source luminosity. Our results
are consistent with an obscuration of the inner disk region by a
partially ionized outer disk.  The size of the corona, derived from
the dip ingress times, was found to be $\sim10^9$ cm. The corona size
did not correlate with the coronal temperature, but seemed to increase
when \Mdot~ also increased. We discuss our findings in the context of
an evaporated accretion disk corona model and an ADAF-type model.

\end{abstract}




\section{Introduction}

The low-mass X-ray binary (LMXB) \xb1916 is the most compact X-ray
binary which exhibit intensity reductions, or dips.  It is generally
accepted that the dips are due to the obscuration of the central
object in a large inclination system by a bulge in the outer accretion
disk \citep{white82a}.  \xb1916 was the first dipping source
discovered, and has an X-ray period of $\sim3000$ seconds
\citep{walter82,white82a}.  Since its discovery, \xb1916 has been the
subject of numerous investigations.  The dip spectra from {\it Exosat}
\citep{smale88} and {\it Ginga} \citep{smale92,yoshida95} observations
were best fit with a two component model. During the dips, one
component was absorbed and the other component decreased in
normalization. More recent observations with {\it ROSAT}
\citep{morley99}, {\it ASCA} \citep{church97}, and {\it Beppo-SAX}
\citep{church98} found the deep dip spectra could be fit with a
partially covered blackbody plus power-law model. In this model, the
blackbody component is heavily absorbed but the power-law
component is only partially absorbed. The normalization remains the
same between dip and non-dip spectra.

The partial covering model implies an emission geometry where a
compact blackbody is surrounded by an extended corona. This picture is
also supported by {\it RXTE} observations of the non-dip emission from
\xb1916 \citep{bloser00a}.  However, there are other dipping LMXBs
which clearly show different components. The eclipse spectrum of EXO
0748-676 shows a residual thermal emission lines from a low
temperature corona while the harder component is heavily obscured
\citep{bonnet00}. The dip spectrum from X 1658-298 shows the blackbody
component with a smaller absorption than the Comptonizing component
\citep{oosterbroek01}. It is puzzling that effects of the outer disk
obscuration of the inner disk region can vary greatly, even in a
relatively small range of inclination angles.

The purpose of this investigation is to study the relationship between
mass accretion rate and spectral and timing behavior of the dips in
\xb1916 using {\it RXTE} data.  LMXBs, in general, show various
spectral states or changes in luminosity.  The change in luminosity
has been typically used as a measure of the mass accretion on the
compact object \citep{mitsuda84}. However, for atoll sources like
\xb1916, it is also generally accepted that the motion within a
color-color diagram (CCD) is a function of mass accretion
\citep{hasinger89}. Several recent observations show a strong
correlation between atoll LMXB spectral parameters and motion within a
CCD \citep{mendez99, bloser00b, gierlinski02a, muno02}. An earlier
analysis of the non-dip spectral shape from \xb1916 also shows a
correlated motion along its CCD which implies variation in \hbox{$\dot
{M}$} \citep{bloser00a}.

\section{Observations and Analysis}

\subsection{RXTE Observation and Data Reduction}

XB 1916-053 was monitored by PCA and HEXTE instrument aboard {\it
Rossi X-ray Timing Explorer} (RXTE) \citep{bradt93} from 1996 February
to 1996 October at roughly one month intervals. An additional 10
observations were performed on consecutive days from 1996 May 14 to
May 23. A detailed log of the observations can be found in
\cite{boirin00}. The analysis of the persistent emission, excluding
dips and bursts, can be found in \cite{bloser00a}.

The gain on the PCA detector has been changed on five occasions since
mission start. Two of our observations were done during PCA Gain Epoch
1 and the remainder during Gain Epoch 3. Since each Epoch requires its
own background models and response matrices, we could not confidently
relate the spectral parameters from one Epoch to another (see Bloser
et al. 2002a). Therefore, we excluded 1996 February 10 and 1996 March
13 data which were taken during PCA Gain Epoch 1 from further
analysis.

For the remaining observations, we created two Good Time Intervals
(GTI). The non-dip GTI excluded bursts, and primary and secondary
dips. The primary dip ephemeris was taken from \cite{chou01}. For
each dip, we averaged the 2-30 keV PCA Standard-2 count rate to
establish a baseline. The portion of the dip where the count rate was
within 10\% of the baseline was accumulated in the dip GTI.
Due to Earth occultation coincident with the dips, we did not record
any dip data from the 1996 May 5 observation. We also did not record
any dip data from 1996 May 15 observation because the dips were not
seen.  Once the GTIs were defined, the non-dip and the deep dip data
were reduced using the standard {\it RXTE} analysis tools in FTOOLS
5.1. The HEXTE spectral data were deadtime corrected and extracted
from the standard archive mode data. The PCA spectral data were
extracted from the Standard-2 data, which has 128 energy channels
between 2 and 100 keV with 16 second time resolution. The PCA response
matrices were generated with {\it pcarsp}. Spectra for each PCU and
HEXTE cluster were reduced separately and combined within XSPEC
v11.1.0.

\subsection{Color-Color Diagram and Lightcurve}

\begin{figure}
\plotone{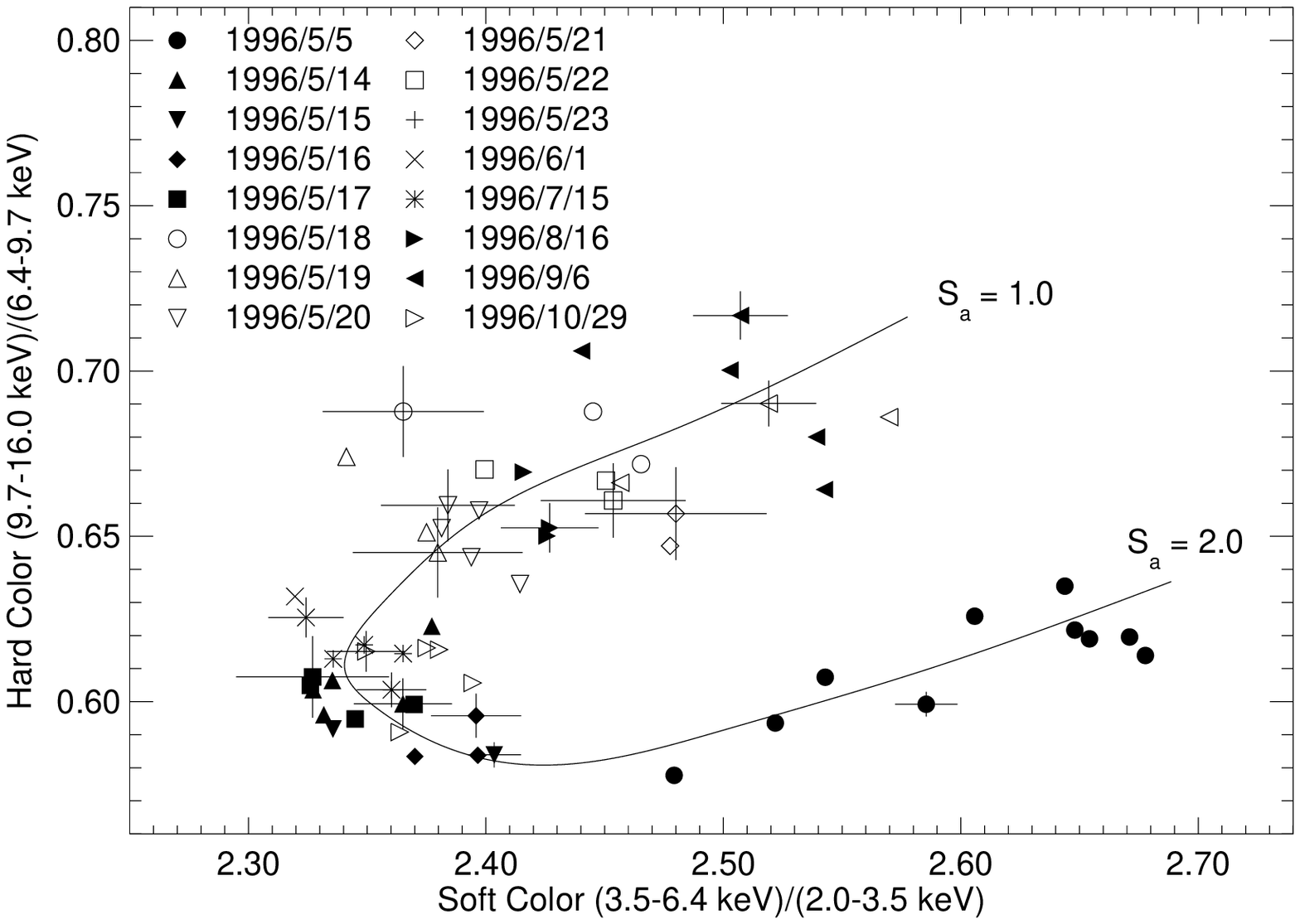}
\caption{Color-color diagram of XB 1916-053. Each point represents a
segment of the persistent emission with a typical length of 1200
s. Various symbols denote the different observation date. For clarity,
only one set of error bars is displayed per date. \label{fig-ccd}}  
\end{figure}

\cite{bloser00a} analyzed the non-dip emission by dividing the entire
RXTE data into 70 segments of typical length $\sim1200$ s. By plotting
each data segment as a point on a CCD, they found that \xb1916
traversed the lower and upper banana branches. Following
\cite{mendez99}, the CCD track was fit to a spline and each data
segment was parameterized with an \sa value on distance along the CCD
location from the upper end of the CCD track for which S$_a = 1.0$ was
arbitrarily assigned. Here we are interested in the accretion rate, or
the \sa parameter, for each observation date. In Figure \ref{fig-ccd},
we replot the CCD using a different symbol for each observation
date. The soft and hard colors are defined as the ratios of
background-subtracted PCA count rates in the bands 3.5-6.4 keV and
2.0-3.5 keV, and 9.7-16.0 keV and 6.4-9.7 keV, respectively. In Table
\ref{tbl-ingress1}, we list an average \sa value for each observation
date using the \sa values from the data segments \citep{bloser00a}.

\begin{figure}
\plotone{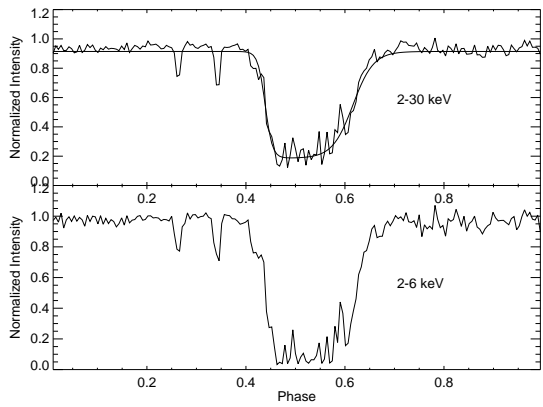}
\caption{Folded lightcurves from May 14 1996 observation in the 2-30 keV 
and 2-6 keV energy bands.  The 2-30 keV band lightcurve shows the
best fit polynomial curve.  \label{fig-lightcurve}}   
\end{figure}

Throughout the monitoring campaign, the ingress/egress times and the
dip durations varied considerably. To quantitatively assess the
relationship between spectral parameters and dip profile, we removed
any bursts during or near dips and folded the Standard-2 data at the
X-ray period of 3000.27 seconds \citep{chou01} for each observation
date. A typical background subtracted lightcurves in the energy ranges
2-30 keV and 2-6 keV are shown in Figure \ref{fig-lightcurve}. In the
observations when the additional dip blackbody absorption is required
(see sec 2.2), the 2-6 keV intensity reduction was nearly 100\% in the
dips. In the broadband 2-30 keV folded lightcurve, the ingress and
egress times were defined as the time to go from 10\% below the
non-dip count rate to 10\% above the dip count rate. The dip duration
was defined as the full-width at half-minimum. The dip profiles were
usually not symmetric so we fitted the ingress and egress lightcurves
separately with a polynomial, and considered the shorter of the two
times as an upper limit measure of the corona size (see sec. 3.2). For
clarity, we hereafter refer to the shorter of the two times as the
ingress time. The results of dip profile fitting are listed in Table
\ref{tbl-ingress1}. The ingress and duration times were found to vary
by a factor of three and factor of thirty respectively throughout our
observations.

\begin{figure}
\plotone{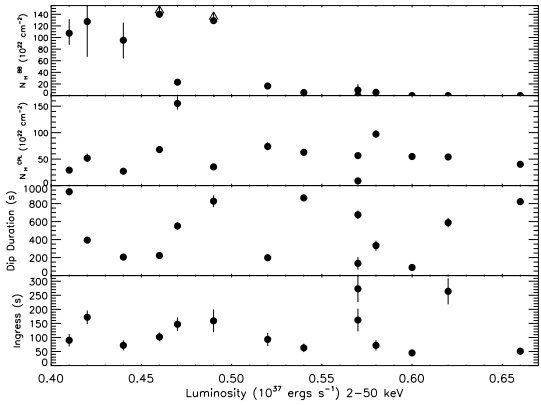}
\caption{Variation of dip blackbody and cut-off power law absorptions,
and dip ingress and duration with luminosity ($2-50$ keV). Error bars
are $1~\sigma$ for one interesting parameter. \label{fig-lum}} 
\end{figure}

In the bottom two panels of Figure \ref{fig-sa} and Figure
\ref{fig-lum} we show the dip duration and ingress time of each
observation as a function of \sa and broadband luminosity (see
sec. 2.3).

\subsection{Spectral Fitting}

For each observation date, single non-dip emission spectrum was
extracted from the combined PCA and HEXTE data sets.  Based on the
spectral fits to the archived Crab data, we used only the data between
2.5 to 20.0 keV from PCUs 0 \& 1. A systematic error of 1\% was added
to all the channels using {\it grppha}. HEXTE data was selected from
20.0 keV to 50.0 keV.  \cite{bloser00a} found that the non-dip
emission could be fit with either a cut-off power law plus a blackbody
model (CPL+BB) or a Comptonization plus a blackbody
(CompTT+BB). Fitting with a disk blackbody plus a cut-off power law
(Disk+CPL) was reported to give unreasonable fits. We too found the
CPL+BB model gave good fits to the data, and could not fit the data
with the Disk+CPL model. We also tried the CompTT+BB model. Although
CompTT+BB gave acceptable fits, the error bars were very large, and we
decided to use only the CPL+BB model for our analysis.  Due to the
PCA's insensitivity below 2.5 keV, the hydrogen column density \nh was
not allowed to vary. Taking an average value from several previous
X-ray observations \citep{smale88, smale92, yoshida95, morley99}, we
fixed \nh $= 2.0 \times 10^{21}$ cm$^{-2}$. We list the best-fit
parameters for the persistent emission in Table \ref{tbl-spectra}.

We extracted the dip PCA spectra in a similar manner as the non-dip
spectra using the dip GTI. However, in many cases the HEXTE count rate
was too low to effectively constrain the dip spectra. Since most of
the spectral differences between non-dip and dip data occurred at
$<20$ keV, we decided to use only the PCA data for the dip analysis.

X-ray dips are almost certainly due to obscuration of the central
region by the outer accretion disk, and thus the non-dip emission
model must also give acceptable fits to the dip emission with only the
\nh allowed to vary. Initially, we tried a very crude model of
increasing the absorption to the persistent emission model. This
consistently gave unacceptable fits. Next, we allowed the absorption
to vary independently between the BB and CPL components. The BB and
CPL normalizations were fixed to the non-dip model. This gave
acceptable fits to all data and the results are reported in Table
\ref{tbl-spectra}.

We proceeded to examine the relationship between the dip spectra and 
the non-dip spectral shape (as parameterized by \sa on the CCD). There
is growing evidence that the motion within the CCD, and not
necessarily luminosity, is the best indication of mass accretion
\hbox{$\dot {M}$} 
in LMXBs \citep{hasinger89,vanderklis01,gierlinski02a}. 
In the top two panels of Figure \ref{fig-sa} we show the BB and CPL
absorption parameters as a function of S$_a$. 
The BB absorption is modest ($<25 \times10^{22}$
cm$^{-2}$) when \sa is smaller than 1.12 or greater than 1.40. In
the intermediate range of S$_a$, the BB absorption fluctuates
greatly. The CPL absorption is relatively constant at $\sim50 - 60
\times10^{22}$ cm$^{-2}$, with occasional increase to $\sim100
\times10^{22}$ cm$^{-2}$.

\begin{figure}
\plotone{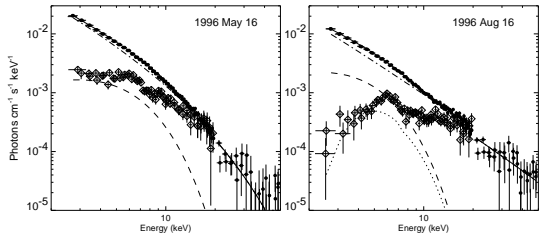}
\caption{Examples of unfolded non-dip and dip spectra. The dash-dot
line is the non-dip CPL model, and the dashed line is the non-dip BB
model. On the left is the 1996 May 16 observation with little
blackbody absorption during dips. On the right is the 1996 Aug 16
observation with large blackbody absorption. The dotted line in the
Aug 16 plot shows the dip BB model. The dip CPL model is not shown
for clarity. \label{fig-spec}} 
\end{figure}

We also compared the dip spectra against the non-dip luminosity.
In the top two panels of Figure \ref{fig-lum} we show the BB and CPL
absorption parameters with respect to the broadband $2-50$ keV
luminosity. The BB absorption 
is $\sim100\times10^{22}$ cm$^{-2}$ or greater at low luminosities.
At $\sim0.5\times 10^{37}$ ergs s$^{-1}$, the additional BB absorption
suddenly decreases to nearly zero and remains very small as the
luminosity increases. Meanwhile, the CPL absorption again remains
relatively constant regardless of the BB absorption. This variation in
BB absorption is seen clearly when comparing the spectra from two
observations. In Figure \ref{fig-spec}, we show an example of a large
BB absorption dip spectrum versus a small BB absorption dip
spectrum. It appears that the attenuation of the $<6$ keV photons is
not nearly as large in the more luminous May 16th observation as in
the Aug 16th observation. The best fit model for the May 16th data
requires no additional BB absorption.

We checked to see whether our results were artifacts of the fitting
procedure. First, we fixed the blackbody absorption to
$1000\times10^{22}$ cm$^{-2}$, essentially eliminating the BB
component, while letting the CPL absorption to vary. This gave
unacceptable fits in all cases where the luminosity was greater than
$0.5\times 10^{37}$ ergs s$^{-1}$. Only May 20 and May 22 observations
gave acceptable fits with no BB component. Second, we removed the
restriction of fitting to the non-dip model and allowed all BB+CPL
parameters to vary. In the majority of the data, the BB temperature,
CPL photon index, and CPL cut-off energy was within $1\sigma$ of the
persistent model. More importantly, the BB absorption was still small
for the bright observations and remained large for the dim
observations.

\section{Summary and Discussion}

We present results of a long term {\it RXTE} monitoring of x-ray dips
in \xb1916. The spectral changes during dips, and the dip
ingress/egress and duration times were examined. The persistent
emission is best fitted by blackbody and cut-off power law
components. During the dips, the flux reduction is greatest in the
2-10 keV band, but extends up to 20 keV. The deep dip spectra are best
described by the non-dip spectral model, usually with additional
absorption for the blackbody and the cut-off power law components.  On
several observations, however, there is no increase in the blackbody
absorption from non-dip to dip spectra. In one observation, the dips
are not seen at all. This is in direct contrast to the {\it Beppo-SAX}
observation which found a strong absorption for the blackbody
component. Our analysis of the 2-30 keV band lightcurves show time
varying dips with complex structure.

\subsection{Dip Absorption}

X-ray dips are most likely due to absorbing material in the outer
accretion disk. One model for the dips proposed by Frank, King \&
Lasota (1987, hereafter FKL) considers the ballistic trajectory of the
accreting material passing above and below the disk. The accretion
stream circularizes close to the central source after shocking several
times \citep{lubow75}, and the ionization instability \citep{krolik81}
creates a bulge consisting of dense cool clouds among hotter
lower-density medium. For the orbital period of \xb1916 and assuming a
neutron star mass of 1.4 M\sun, FKL model predicts the radius of the
circularizing ring to be $r_h=4.1\times10^9$ cm, with the vertical
height of the bulge fixed at $\sim0.4 r_h$. The column density of an
individual cloud is expected to be $\sim3\times10^{24}$ cm$^{-2}$,
consistent with our lower limits for the lower luminosity
observations.

The small absorption of the blackbody component seen in \xb1916 may be
due to partial ionization of the absorbing material.  Partial
ionization of the absorber have been proposed to explain the low
energy excess seen in dips from several high mass X-ray binaries and
black hole candidates in outburst \citep{marshall93, kuulkers97}.
Although \xb1916 is a relatively dim source, the system is much more
compact and thus its outer disk could be ionized.  At the
circularization radius, the ionization parameter is $\xi=L/nr_h^2$,
where $L$ is the source luminosity and the $n$ is the number density.
We can estimate the number density as $n = {\rm N}_{\rm H}/ \Delta r$,
where $\Delta r$ is the thickness of the absorbing material. With
\nh$\sim100$ cm$^{-2}$ and letting $\Delta r \sim 0.1 r_h$, the number
density is $n\sim2\times10^{15}$ cm$^{-3}$. At the highest observed
luminosity, the ionization parameter is $\xi>150$ erg cm
s$^{-1}$. Near this value of $\xi$ and at a density of
$n\sim2\times10^{15}$ cm$^{-3}$, the low-Z elements in the bulge are
significantly ionized \citep{kallman01}.

Furthermore, the degree of ionization could be even higher in regions
of low gas density in the bulge. The cold clouds, which initially
condense out at the impact point, are expected to fall back toward the
disk within $\sim0.5$ orbit (FKL). This may result in a lower density
of gas near the top of the bulge, similar to the density variation
observed from 4U 1822-37 \citep{white82b}. Meanwhile, the near
constant CPL absorption is likely due to the line of sight to the
extended corona passing through both high and low density regions.

\subsection{Physical extent of the corona}

The CPL emission seen in many LMXBs are believed to originate in a hot
optically thin gas nearby or surrounding the accretion disk.  As the
emission region or some fraction is occulted by the outer disk, we can
use the ingress time to constrain the size of the emitting
region or the absorber.  If the angular extent of the absorber is
larger than the emitting region, the size of the hot gas region is
$d=2 \pi \Delta T r_h/P$, where $\Delta T$ is the ingress time and $P$
is the period. On the other hand, if the absorber has a smaller angular
extent than the emitting region, the equation above gives the size of
the absorber. From the observed ingress times, the extent of either the
CPL emission region or the absorber varies from $d= 3.9 - 23.2 \times
10^{8}$ cm.  To discern between the geometries, we examined the
lightcurves and the spectra from all the observations.
We find that in observations where the
additional dip BB absorption is required, the 2-6 keV band dip intensity
reduction is nearly 100\%, implying that the CPL emission region is
entirely covered by the absorber. We also find the dip CPL absorption
to be nearly constant in all our observations, which suggests that
there are no significant differences in the angular sizes of the
emitting region or the absorber from one observation to another. We
therefore conclude that the ingress time constrains the size of the
CPL emission region.

The possible mechanism for producing the hot gas can be divided
into two broad categories. First possibility is the evaporation of the
disk due to several forms of internal heating \citep{liang77,
paczynski78} or external irradiation \citep{shakura73}, and the
formation of an accretion disk corona (ADC). The extent of an ADC in
hydrostatic equilibrium is governed by the temperature of the gas. The
second possibility is an optically thin advection-dominated accretion
flow (ADAF) type model. In such a case, the hot gas would reside
within the optically thick accretion disk and the extent is likely a
function of the mass accretion rate.  With our observations, we can
test both possibilities by comparing the CPL emission region size, as
inferred from the ingress time, to temperature or mass accretion.

\begin{figure}
\plotone{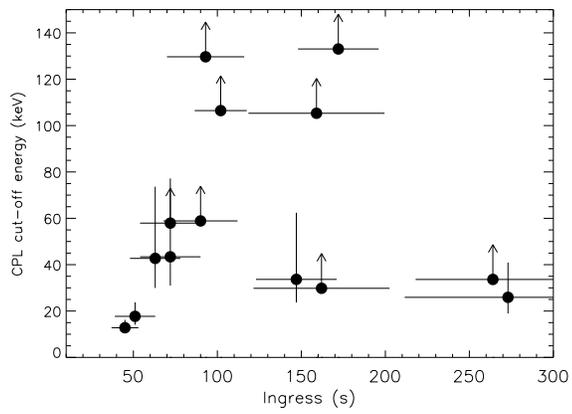}
\caption{Comparison of CPL cut-off energy with ingress
time. \label{fig-ing}}  
\end{figure}

The extent of an ADC surrounding a compact object can be estimated
from the relation $R = 1\times10^{10} M/T_8$ where M is the mass of
the neutron star in M\sun, and the T is coronal temperature in $10^8$
K \citep{begelman83}. Inferring the coronal temperature from the
power-law cutoff energy \citep{haardt93}, we estimate the maximum
radius of an optically thin and an optically thick corona at $kT=12$
keV to be $1\times 10^{10}$ cm and $4\times 10^{10}$ cm,
respectively. According to FKL, a corona cannot extend beyond the
circularization radius. Thus, in \xb1916 we expect the radius to
increase as the temperature decreases, but become bound at the
circularization radius ($4.1\times10^9$ cm) at the lowest
temperatures.  However, when we compare the CPL emission region size,
as parameterized by the ingress time, to the CPL cut-off energy
(Fig. \ref{fig-ing}), we find that the ingress time does not
necessarily increase as the cut-off energy decreases. In fact, the
smallest ingress time is associated with the lowest cut-off energy. We
note however, that the size of the corona could be smaller if the
heating is due mostly to accretion \citep{witt97} or if an accretion
disk wind develops at a much smaller radius \citep{murray95}.

\begin{figure}
\plotone{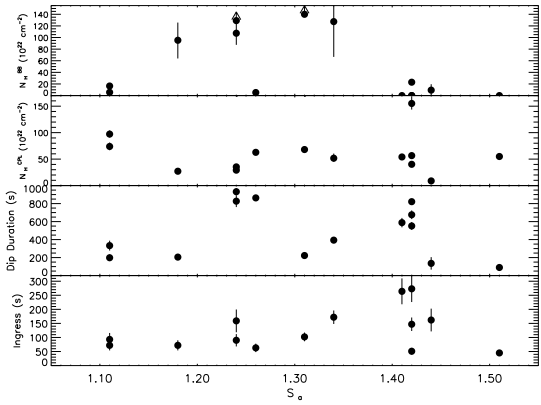}
\caption{Variation of dip blackbody and cut-off power law absorption,
and dip ingress and duration with accretion rate, parameterized by
S$_a$. Error bars are $1~\sigma$ for one interesting
parameter. \label{fig-sa}}  
\end{figure}

Following Shapiro, Lightman \& Eardley solution (1973), during period
of low mass accretion in black hole systems, the inner part of the
disk could be replaced by a hot optically thin flow, such as an ADAF
\citep{narayan94}. When ADAF-type models are applied to black holes,
the truncation radius, i.e. the extent of the hot gas, is highly
uncertain but in all cases decreases with increasing \hbox{$\dot {M}$}
\citep{abramowicz95, liu99, rozanska00}.  The application of an ADAF
model to a neutron star system is more difficult, however, due to the
presence of a boundary layer \citep{narayan97, blandford99,
medvedev01}.  In our observations, the size of the CPL emission, as
parameterized by the ingress time, does not become smaller with
increasing S$_a$, corresponding to increasing \hbox{$\dot {M}$}
(Fig. \ref{fig-sa}). The ingress times remain roughly constant for low
values of S$_a$, and it actually increases to the largest value around
\sa$\sim1.4$.  This may indicate that the observed accretion rate from
\xb1916 is too large, and an ADAF-type flow in a neutron star system
can only occur during periods of very low accretion (e.g. an island
state)\citep{gierlinski02b}.  Another possibility is that the luminous
portion of the ADAF may be much smaller than the truncation radius
\citep{medvedev01}, which will make the correlation between \sa and
ingress time more uncertain.  Finally, other observational evidence
suggests that even if an ADAF like flow exists, its contribution to
the CPL spectrum is minimal \citep{menou02}. Although it is
interesting to note that an apparent decrease in the emission region
size at \sa$>1.45$ (Fig. \ref{fig-sa}) coincides roughly with the
sudden increase in the non-dip $L_{BB}/L_{CPL}$ luminosity ratio
\citep{bloser00b}. Such a scenario might be expected when the thermal
photons from the boundary cools the ADAF-flow and causes it to
collapse.

From our {\it RXTE} observations of \xb1916, we cannot say with
certainty that the CPL emission originates in the ADC or the ADAF-type
flow. We find that the extent of the hot gas can vary on days
timescale, but the size does not correlate with temperature or mass
accretion, as expected in simple ADC or ADAF models. What might be
more likely is the coexistence of both types of hot plasma around the
neutron star \citep{esin97, rozanska00}. In such a case, the
truncation radius of the disk, and thus the extent of the
ADAF-flow, may be better constrained by correlating the low frequency
quasi-periodic oscillations with the dip ingress times. This test may
be better suited for black hole systems where the flow is not affected
by the neutron star boundary layer.

\acknowledgments

TN acknowledges support from the College of the Holy Cross. PFB is a
National Research Council Research Associate at NASA/Goddard Space
Flight Center. This research has made use of data obtained through the
High Energy Astrophysics Science Archive Research Center Online
Service, provided by the NASA/Goddard Space Flight Center.

\clearpage 

\clearpage 

\clearpage 

\clearpage 

\clearpage


\clearpage


\clearpage

\clearpage

\begin{deluxetable}{crrrr}
\tabletypesize{\scriptsize}
\tablecaption{Dip Ingress/Egress and Duration. \label{tbl-ingress1}} 
\tablewidth{0pt}
\tablehead{
\colhead{Obs Date} & 
\colhead{$S_a$} & 
\colhead{L\tablenotemark{a}} & 
\colhead{Ingress (s)} & 
\colhead{Duration (s)} \\
}

\startdata
May 14 1996 & 1.42 & 0.47 & $147\pm24$ & $551\pm43$ \\
May 16 1996 & 1.51 & 0.60 & $45\pm8$ & $90\pm13$ \\
May 17 1996 & 1.44 & 0.57 & $162\pm41$ & $136\pm69$ \\
May 18 1996 & 1.24 & 0.41 & $90\pm22$ & $932\pm38$ \\
May 19 1996 & 1.34 & 0.42 & $172\pm24$ & $393\pm32$ \\
May 20 1996 & 1.31 & 0.46 & $102\pm16$ & $222\pm28$ \\
May 21 1996 & 1.18 & 0.44 & $72\pm18$ & $205\pm33$ \\
May 22 1996 & 1.24 & 0.49 & $159\pm41$ & $826\pm65$ \\
May 23 1996 & 1.41 & 0.62 & $264\pm46$ & $588\pm52$ \\
June 1 1996 & 1.42 & 0.66 & $51\pm12$ & $821\pm28$ \\
July 15 1996 & 1.26 & 0.54 & $63\pm15$ & $863\pm33$ \\
Aug  16 1996 & 1.11 & 0.52 & $93\pm23$ & $198\pm37$ \\
Sept 6  1996 & 1.42 & 0.57 & $273\pm62$ & $676\pm48$ \\
Oct 29  1996 & 1.11 & 0.58 & $72\pm18$ & $333\pm55$ \\

\enddata

\tablecomments{Errors are $1\sigma$ for one interesting parameter.}

\tablenotetext{a}{Total 2-50 keV luminosity, times $10^{37}$ ergs
s$^{-1}$ for a distance of 9.3 kpc.}

\end{deluxetable}

\clearpage

%% Text for table notes should follow after the \enddata but before
%% the \end{deluxetable}. Make sure there is at least one \tablenotemark
%% in the table for each \tablenotetext.
\begin{deluxetable}{crrrrr|rrr}
\tabletypesize{\scriptsize}
\tablecaption{Persistent Emission and Dip Spectral Fits of
\xb1916. \label{tbl-spectra}} 
\tablewidth{0pt}
\tablehead{
\colhead{} & \colhead{} & \multicolumn{4}{c}{Persistent Emission} &
\multicolumn{3}{c}{Dip Emission} \\
\colhead{Obs Date} &
\colhead{$S_a$} & 
\colhead{$kT_{BB}$\tablenotemark{a}} &
\colhead{$\alpha$\tablenotemark{b}} & 
\colhead{$E_c$\tablenotemark{c}} & 
\colhead{L\tablenotemark{d}} & 
\colhead{N$_{\rm H}$(BB)\tablenotemark{e}} &
\colhead{N$_{\rm H}$(CPL)\tablenotemark{f}} &
\colhead{$\chi^2_{\nu}$}
}

\startdata

May 14 1996 & 1.42 & $1.48\pm0.06$ & $1.89\pm0.09$ &
$33.66_{-9.05}^{+28.74}$ & 0.47 & $23.1\pm3.1$ & $155.1\pm11.4$ & 0.89 \\
May 16 1996 & 1.51 & $1.70\pm0.06$ & $1.64\pm0.08$ &
$12.80_{-2.22}^{+3.42}$ & 0.60 & $0+0.220$ & $55.0\pm5.3$ & 1.07 \\
May 17 1996 & 1.44 & $1.58\pm0.05$ & $1.99\pm0.10$ &
$>29.8$ & 0.57 & $9.41\pm9.50$ & $8.85\pm1.7$ & 0.94 \\
May 18 1996 & 1.24 & $1.39\pm0.05$ & $1.89\pm0.03$ &
$>58.9$ & 0.41 & $107.5\pm22.5$ & $29.1\pm1.5$ & 1.12 \\
May 19 1996 & 1.34 & $1.50\pm0.11$ & $1.83\pm0.21$ &
$>133.0$ & 0.42 & $>66$ & $51.7\pm5.5$ & 0.61 \\
May 20 1996 & 1.31 & $1.47\pm0.04$ & $1.96\pm0.02$ &
$>106.5$ & 0.46 & $>140$ & $68.1\pm5.8$ & 0.77 \\
May 21 1996 & 1.18 & $1.45\pm0.05$ & $1.92\pm0.10$ &
$>57.9$ & 0.44 & $95.3\pm31.0$ & $27.1\pm3.4$ & 0.43 \\
May 22 1996 & 1.24 & $1.45\pm0.04$ & $1.94\pm0.06$ &
$>105.3$ & 0.49 & $>129$ & $35.3\pm1.1$ & 0.86 \\
May 23 1996 & 1.41 & $1.60\pm0.05$ & $2.05\pm0.13$ &
$>33.6$ & 0.62 & $0+0.5$ & $54.1\pm3.6$ & 1.29 \\
June 1 1996 & 1.42 & $1.56\pm0.02$ & $1.74\pm0.09$ &
$17.72_{-3.65}^{+6.09}$ & 0.66 & $0+0.2$ & $40.2\pm2.1$ & 1.40 \\
July 15 1996 & 1.26 & $1.44\pm0.06$ & $1.82\pm0.08$ &
$42.71_{-12.76}^{+30.94}$ & 0.54 & $5.27\pm1.0$ & $62.8\pm2.7$ & 1.07 \\
Aug  16 1996 & 1.11 & $1.34\pm0.03$ & $1.81\pm0.02$ &
$>129.7$ & 0.52 & $16.50\pm2.0$ & $73.9\pm6.2$ & 0.75 \\
Sept 6  1996 & 1.42 & $1.64\pm0.06$ & $1.87\pm0.11$ &
$25.92_{-6.97}^{+15.01}$ & 0.57 & $0+0.10$ & $56.6\pm2.6$ & 1.02 \\
Oct 29  1996 & 1.11 & $1.40\pm0.06$ & $1.74\pm0.08$ &
$43.40_{-12.42}^{+33.77}$ & 0.58 & $5.66\pm1.2$ & $97.3\pm6.3$ & 0.67 \\

\enddata
\tablecomments{For all persistent emission fits \nh is frozen at
$0.2\times10^{22}$ cm$^{-2}$. Errors are $1\sigma$ for
one interesting parameter.}

\tablenotetext{a}{Blackbody temperature (keV).}
\tablenotetext{b}{Cut-off power law photon index.}
\tablenotetext{c}{Cut-off power law cut-off energy (keV).}
\tablenotetext{d}{Total 2-50 keV luminosity, times $10^{37}$ ergs
s$^{-1}$ for a distance of 9.3 kpc.}
\tablenotetext{e}{\nh, $\times10^{22}$ cm$^{-2}$, for blackbody
component.} 
\tablenotetext{f}{\nh, $\times10^{22}$ cm$^{-2}$, for cut-off power
law component.} 

\end{deluxetable}

\end{document}